# Interchange instability in an accretion disc with a poloidal magnetic field

H.C. Spruit,[1,2] R. Stehle[1,2] and J.C.B. Papaloizou[3]
[1] *Astronomical Institute 'Anton Pannekoek', Kruislaan 403, 1098 SJ Amsterdam, Netherlands*
[2] *Max Planck Institute for Astrophysics, Box 1523, D-85740 Garching, Germany*
[3] *Theoretical Astronomy Unit, Queen Mary and Westfield College, Mile End Road, London E1 4NS*

**ABSTRACT**
We investigate the stability to nonaxisymmetric perturbations of an accretion disc in which a poloidal magnetic field provides part of the radial support against gravity. Interchange instability due to radial gradients in the magnetic field are strongly stabilized by the shear flow in the disc. For smooth field distributions this instability is restricted to discs in which the magnetic energy is comparable to the gravitational energy. An incompressible model for the instability akin to the Boussinesq approximation for convection is given which predicts the behaviour of the instability accurately. Global axisymmetric disturbances are also considered and found to be stable for a certain class of models. The results indicate that accretion discs may be able to support poloidal fields which are strong enough to suppress other forms of magnetic instability. These strong and stable field distributions are likely to be well suited for the magnetic acceleration of jets and winds.

## 1 INTRODUCTION

Magnetic fields in an accretion disc are advected with the accretion flow. The net poloidal flux (flux passing through the disc) can not be changed by small scale processes inside the disc itself. This flux is 'inherited' by the disc from the accreting matter. Compressed by the accretion flow, this flux can build up substantial field strengths in the central regions. At least in the case of protostellar discs, the enviroment in which they form are known to be highly magnetic (Heiles et al. 1993), and the question how, while forming a star, the accretion flow disposes of the 'excess flux' of magnetic field lines threading the disc is a classical problem in star formation (see the review by McKee et al. 1993).

Depending on their strength and radial distribution, poloidal fields can be of the right strength and configuration to produce magnetically accelerated jets and winds (Königl and Ruden 1993, Blandford 1992). Acceleration of these flows works best in strong fields, while their degree of collimation is probably dependent on the radial distribution of the poloidal flux (cf. Spruit, 1994). The strength of the field is limited by the rate at which the field diffuses outward, either by some 'turbulent' process (e.g. Van Ballegooijen, 1989), or by microscopic processes (in the case of protostellar discs). Of particular interest is outward diffusion due to instabilities driven by the magnetic field itself, which is expected to become important especially at high field strength. Such instabilities are expected, in the form of *interchanges* whenever the magnetic forces provide part of the support against gravity. In the same way as in solar prominences (Anzer 1969, Priest 1982), energy is gained in this case by exchange of the magnetized gas providing the support against the less strongly magnetized gas being supported. In this form of instability, neighboring field lines with the plasma attached to them are interchanged, with little or no bending of the field lines. For poloidal fields threading a disk, this means that the unstable displacements take place in the disk plane, with little dependence on the direction perpendicular to the plane. This interchange process is known to be quite effective at transporting mass across field lines in solar prominences, where it can be studied in detail (cf Priest 1982, Schmieder 1989). The process characteristically takes the form of small scale (local) interchanges. It has previously been studied in the context of rotating, magnetized stars by Pitts and Tayler (1985, and references therein).

In the context of accretion discs, interchange instability has been studied by Spruit and Taam (1989, hereafter ST), who concluded that uniform rotation has no effect on the instability condition or the maximum growth rates (see also Lepeltier and Aly, 1995). From this, they concluded that transfer of mass *across* field lines may well be quite effective in the magnetospheres of accreting neutron stars. These re-sults are not applicable to fields in accretion discs, since they hold for uniform rotation only, and do not take into account the strong *shear* of the flow in a disc. Shear is likely to have a strongly stabilizing effect, since it distorts unstable displacements, on the short orbital time scale, into less rapidly growing ones.

In this paper, we study the effect of shear on interchange instability. We first point out the close analogy between this process and 2-dimensional convection, by showing that in an approximation akin to the incompressible (Boussinesq) approximation in convection, the two problems are mathematically identical. By integrating the full time dependent equations for linear disturbances in a local (shearing sheet) model we show that the instability behaves, in the short wavelength limit, as expected from an incompressible model. The effect of shear turns out to be quite strong, such that the instability becomes important only when the magnetic support of the disc becomes comparable to gravity, or when the field strength varies on a short radial length scale.

Other forms of instability of poloidal fields are known, in particular the rotating-shear flow instability studied in the context of discs by Balbus and Hawley (1991). This instability is suppressed already when the field strength is comparable to the gas pressure. An important conclusion is therefore, that poloidal fields larger than this value are *stable* to both the Balbus-Hawley form and to interchanges, up to the point where the field starts contributing significantly to the support of the disc. The strong poloidal fields that are optimal for the production of jets and winds may thus be a natural consequence of the concentration of fields by the accretion process.

A preliminary treatment of the present problem is reported in Lubow and Spruit (1995).

## 2 EQUATIONS

In the thin disc limit, neither the gas pressure nor the magnetic pressure contributes, and only the magnetic curvature force appears in the equation of motion. Since to first approximation the interchanges do not bend field lines and are nearly parallel to the disk plane, we can take the flow field $\boldsymbol{u}$ to be independent of height in the disk. As shown in ST, the MHD equations can then be reduced to two dimensions by integrating vertically across the disk. In the absence of viscous forces, these thin-disc MHD equations are, for motions in the disk plane:

$$\frac{\mathrm{d}\boldsymbol{u}}{\mathrm{d}t} = \frac{[\boldsymbol{B}]B_z}{4\pi\Sigma} + \boldsymbol{g}, \tag{1}$$

$$\partial_t \Sigma + \mathrm{div}(\Sigma \boldsymbol{u}) = 0, \tag{2}$$

$$\partial_t B_z + \mathrm{div}(B_z \boldsymbol{u}) = 0, \tag{3}$$

where $[\boldsymbol{B}]$ denotes the jump in the field vector across the disc and $\boldsymbol{g}$ is the acceleration of gravity. The disc is assumed to be flat and perpendicular to the z-axis and to remain so during perturbations. That is, we do not consider 'buckling modes' or corrugation waves (for more general equations allowing for corrugations see ST). Due to the thin disc assumption, all structure inside the disc has disappeared from view, and the curvature force, integrated across the disc appears as the first term on the RHS in (1). The continuity and induction eqs. (2,3) can be combined into:

$$\mathrm{d}(\Sigma/B_z)/\mathrm{d}t = 0. \tag{4}$$

To complete the equations, a prescription is needed to determine the field configuration above the disc from the values of the normal component $B_z$ on the disc. We assume the field above the disc to be a potential field:

$$\boldsymbol{B} = \boldsymbol{\nabla}\Phi, \qquad \nabla^2 \Phi = 0, \tag{5}$$

with boundary conditions

$$\partial_z \Phi = B_z \quad (z=0), \quad \Phi \to 0 \quad (z \to \infty). \tag{6}$$

In general, one would like to solve for a force-free field above the disc, but this is a much more demanding problem. Instead, we assume here that the Alfvén speed above the disc is sufficiently high that efficient reconnection to a nearly potential field takes place (as is the case in the solar corona, for example).

Conceptually, the motions determined by the equation of motion carry the vertical component of the field around by the induction equation (3); the potential field problem (5) determines from $B_z$ the horizontal components $(B_r, B_\phi)$ at the disc surface. These components then determine the magnetic force in the equation of motion, thus closing the set of equations.

### 2.1 Analogy with 2-D hydrodynamics

The form of the equations has a certain similarity to the ordinary hydrodynamic equations in two dimensions. This is seen by noting that

$$(\boldsymbol{\nabla} \times [\boldsymbol{B}])_z = 2j_z = 0 \tag{7}$$

by the potential field assumption. Hence we may write

$$[\boldsymbol{B}]/4\pi = -\boldsymbol{\nabla}_\mathrm{h} f, \tag{8}$$

where $\boldsymbol{\nabla}_\mathrm{h}$ denotes the gradient taken in the plane of the disc. Writing

$$m \equiv \Sigma/B_z, \tag{9}$$

eqs. (1) and (4) can be written as

$$m\frac{\mathrm{d}\boldsymbol{u}}{\mathrm{d}t} = -\boldsymbol{\nabla}_\mathrm{h} f + m\boldsymbol{g}, \tag{10}$$

$$\mathrm{d}m/\mathrm{d}t = 0. \tag{11}$$

Equation (10) is identical to the Euler equation, with $m$ playing the role of the gas density $\rho$ and $f$ that of the gas pressure $p$. The form of the 'pressure' $f$ is very different from the ordinary hydrodynamic case, however. Whereas in hydrodynamics one can assume, say, a polytropic relation between $p$ and $\rho$ or an ideal gas, $f$ is related to $m$ in a much more complicated way, involving eqs. (11), 3 and the solution of the potential problem (5). Hence the analogy with hydrodynamics is really close only if additional assumptions are made. The problem defined by (1-3), (5) in general has compressive solutions (i.e. $\mathrm{div}\,\boldsymbol{u} \neq 0$), mediated by the 'pressure' $f$ (ST, Tagger et al. 1990). The interchange instabilities studied in this paper, however, are basically incompressive in nature, and it turns out to be illuminating to describe them in an incompressible approximation which filters out the stable compressive waves. This can be done by taking the curl of (10) to eliminate $f$ and adding the approximation $\mathrm{div}\,\boldsymbol{u} = 0$. In this way, the incompressible limit of the magnetic instability problem becomes equivalent to 2-dimensional (z-independent) Bousinesq convection. The incompressible limit is discussed further in section 3.2.

## 2.2 Linearized equations in disc geometry

We write equations (1,3,4) in polar coordinates $(r, \phi)$ and linearize about a basic state $u_r = 0$, $u_\phi = \Omega(r)r$. In the absence of currugations of the disc, and of externally imposed fields, the field is antisymmetric (dipole-like) about the midplane, such that $B_z$ is the same on both sides, while the values of $B_r$ and $B_\phi$ are opposite. Thus we may write

$$[B_r] = 2B_r^+, \qquad [B_\phi] = 2B_\phi^+, \tag{12}$$

where the index $^+$ denotes evaluation at the upper surface of the disc. We denote by a prime $'$ the (Eulerian) perturbations while unprimed quantities refer to the equilibrium state, which satisfies

$$0 = g_{\rm m} - g + \Omega^2 r, \tag{13}$$

where

$$g_{\rm m} = \frac{B_r^+ B_z}{2\pi \Sigma} \tag{14}$$

is the magnetic acceleration and $g = GM/r^2$ the gravity due to the central object. Due to the magnetic support, $\Omega$ differs from its Keplerian value. The shear rate in this equilibrium is

$$S \equiv r \frac{\mathrm{d}\Omega}{\mathrm{d}r} = -\frac{3}{2}\Omega - \frac{g_{\rm m}}{2\Omega}\frac{\mathrm{d}}{\mathrm{d}r}\ln(r^2 g_{\rm m}). \tag{15}$$

This also differs from the Keplerian expression $S_{\rm K} = -3/2\Omega_{\rm K}$. The linearisation of (1,3,4) yields

$$(\partial_t + \Omega\partial_\phi)u'_r - 2\Omega u'_\phi = \frac{B_r^{'+} B_z}{2\pi\Sigma} + \frac{B_r^+}{2\pi}\left(\frac{B_z}{\Sigma}\right)', \tag{16}$$

$$(\partial_t + \Omega\partial_\phi)u'_\phi + u'_r \frac{1}{r}\frac{\mathrm{d}}{\mathrm{d}r}(\Omega r^2) = \frac{B_\phi^{'+} B_z}{2\pi\Sigma}, \tag{17}$$

$$(\partial_t + \Omega\partial_\phi)B'_z = -\mathrm{div}(B_z u'), \tag{18}$$

$$(\partial_t + \Omega\partial_\phi)\left(\frac{\Sigma}{B_z}\right)' + u'_r \frac{\mathrm{d}}{\mathrm{d}r}\left(\frac{\Sigma}{B_z}\right) = 0. \tag{19}$$

## 2.3 Shearing sheet model

An approximate model for a differentially rotating disc is the shearing sheet (Goldreich and Lynden-Bell, 1965). In this model local cartesian coordinates (x,y) in the radial and azimuthal directions are used around a point $r = r_0$, in a frame rotating at the rate $\Omega_0 = \Omega(r_0)$, while the shear is approximated as constant, i.e.

$$r = r_0 + x, \qquad 1/r_0 \partial_\phi \to \partial_y, \qquad \Omega - \Omega_0 = Sx/r_0. \tag{20}$$

This model describes a small section of a disc near $r = r_0$, such that the curvature of the rotating flow can be neglected, and the $r$-dependence of the equilibrium quantities is assumed to be linear. In this approximation, the equations become

$$(\partial_t + Sx\partial_y)u'_x - 2\Omega_0 u'_y = \frac{B_x^{'+} B_{z0}}{2\pi\Sigma_0} + \frac{B_{x0}^+}{2\pi}\left(\frac{B_z}{\Sigma}\right)', \tag{21}$$

$$(\partial_t + Sx\partial_y)u'_y + (2\Omega_0 + S)u'_x = \frac{B_y^{'+} B_{z0}}{2\pi\Sigma_0}, \tag{22}$$

$$(\partial_t + Sx\partial_y)B'_z = -\mathrm{div}(B_z u'), \tag{23}$$

$$(\partial_t + Sx\partial_y)\left(\frac{\Sigma}{B_z}\right)' + u'_r\left(\frac{\mathrm{d}}{\mathrm{d}r}\frac{\Sigma}{B_z}\right)_0 = 0, \tag{24}$$

where the index $_0$ denotes evaluation at $r = r_0$.

An approximation is needed to simplify the potential problem (5). As in ST, we use a short-wavelength approximation, in which the length scale of the perturbation is assumed to be small compared with the length scale of the background gradients. Thus, if the wavenumber is $\boldsymbol{k} = (k_x, k_y)$, the field perturbation has the form

$$\boldsymbol{B}'(x,y,z,t) = \boldsymbol{b}(t)\exp(ik_x x + ik_y y - k|z|), \tag{25}$$

where

$$k = (k_x^2 + k_y^2)^{1/2}, \tag{26}$$

$$b_x = -i\frac{k_x}{k}b_z \,\mathrm{sgn}(z), \tag{27}$$

$$b_y = -i\frac{k_y}{k}b_z \,\mathrm{sgn}(z). \tag{28}$$

We transform to a shearing coordinate frame $x' = x$, $y' = y - Sxt$, $z' = z$, $t' = t$. Hence $\partial_t + Sx\partial_y = \partial'_t$, $\partial_x = \partial'_x - st'\partial'_y$. In these coordinates eqs. (21-24) have solutions of the form

$$B_z \boldsymbol{u} = B_{z0}\boldsymbol{U}(t')\exp(ik_x x' + ik_y y'). \tag{29}$$

The linearized equations then take the form

$$\partial_{t'} U_x = \frac{B_{z0}^2}{2\pi\Sigma_0}\frac{k_x}{k}\beta + F + 2\Omega_0 U_y, \tag{30}$$

$$\partial_{t'} U_y = \frac{B_{z0}^2}{2\pi\Sigma_0}\frac{k_y}{k}\beta - (2\Omega_0 + S)U_x, \tag{31}$$

$$\partial_{t'}\beta = -\boldsymbol{k}\cdot\boldsymbol{U} \tag{32}$$

$$\partial_{t'} F = U_x \frac{B_{x0}^+ B_{z0}}{2\pi\Sigma_0}\frac{\mathrm{d}}{\mathrm{d}r}\ln\left(\frac{\Sigma}{B_z}\right)_0, \tag{33}$$

where

$$\beta = -ib_z/B_z, \qquad F = \frac{B_{x0}^+}{2\pi}\left(\frac{B_z}{\Sigma}\right)'. \tag{34}$$

The wave vector is now time dependent due to the shearing coordinate frame:

$$\boldsymbol{k}(t) = (-Stk_y, k_y), \tag{35}$$

which implies $k_x = 0$ at $t = 0$. As long as $S \neq 0$, other values of $k_x(0)$ are equivalent to a differerent choice of the origin $t = 0$. In the uniformly rotating case $S = 0$, $k_x$ is a constant (generally nonzero) parameter.

The terms involving $\beta$ in the equations of motion represent the restoring force due compression of the field lines. Since this force is mediated by the external potential field, through the curvature force rather than by the magnetic pressure, these terms depend in a different way on the wavenumber than the normal pressure terms. As a consequence, the phase speed of waves mediated by these curvature forces is proportional to $k^{1/2}$.

By introducing shearing coordinates, we have removed the linear dependence of the shear flow on $x$, trading it in for an explicit time dependence of the equations. Since the waves we study are not exponentially growing, their detailed time dependence has to be taken into account anyway (cf.

also the discussion in Ryu and Goodman, 1992). Note that solutions of the form chosen (in shearing coordinates) require that the medium can be regarded as unbounded, or that periodic boundary conditions can be adopted in the direction of the shear ($x$-direction)

From now on, we drop the subscripts $_0$ on the quantities evaluated at $r_0$. The solutions of eqs (30-35) depend on only three independent parameters. This is seen by writing them in nondimensional form. The system has a characteristic length scale:

$$L = \frac{B_z^2}{2\pi\Sigma\Omega^2}. \tag{36}$$

The ratio $L/r$ can be written as

$$L/r = \frac{B_z}{B_x^+} \frac{g_m}{\Omega^2 r} \tag{37}$$

which measures the strength of the magnetic relative to the rotational acceleration in the equilibrium state. Writing

$$\boldsymbol{k} = \boldsymbol{K}/L, \qquad t' = \tau/\Omega, \tag{38}$$

$$\boldsymbol{U} = \Omega L \tilde{\boldsymbol{U}}, \qquad F = \Omega^2 L \tilde{F}, \tag{39}$$

the equations become, after dropping the tilde's:

$$\partial_\tau U_x = \frac{K_x}{K}\beta + F + 2U_y, \tag{40}$$

$$\partial_\tau U_y = \frac{K_y}{K}\beta - (2+s)U_x, \tag{41}$$

$$\partial_\tau \beta = -\boldsymbol{K} \cdot \boldsymbol{U}, \tag{42}$$

$$\partial_\tau F = -U_x a, \tag{43}$$

where

$$a = N_m^2/\Omega^2 = -\frac{g_m}{\Omega^2}\frac{d}{dr}\ln\frac{\Sigma}{B_z}, \tag{44}$$

$$s = S/\Omega. \tag{45}$$

The parameters of the problem are thus $K_y$, $a$ and $s$. The value of $K_x(t)$ is fixed, apart from an arbitrary zero point in time, by the evolution equation (35). In (44) we have introduced, for future reference, the *magnetic buoyancy frequency* $N_m$. It is the frequency of interchange displacements in the case of a stable magnetic gradient, analogous to the buoyancy frequency in a convectively stable stratification.

For the shearing sheet model to be applicable, the wavelength of the unstable flow must be shorter than the length scale on which the unstable gradient is present, at least during the time that growth takes place. Thus we must require that

$$k \gtrsim \frac{d}{dr}\ln\frac{\Sigma}{B_z}. \tag{46}$$

In terms of the dimensionless wavenumber $K$, this implies, approximately,

$$K \gtrsim |a|. \tag{47}$$

## 3 SOLUTIONS

### 3.1 Uniform rotation

The uniformly rotating case has been studied before by ST (see also the discussion in Lubow and Spruit, 1995). Since the wavenumber does not depend on time in this case, eqs. (40-43) have solutions of the form $\exp(\sigma\tau)$, with dispersion relation

$$\left(\frac{\sigma^2}{K}\right)^2 + \frac{\sigma^2}{K}(\frac{4+a}{K}+1) + \frac{a}{K}\left(\frac{K_y}{K}\right)^2 = 0. \tag{48}$$

From the sign of the last term it follows that the condition for instability is $a < 0$ (except in the axisymmetric case $K_y = 0$, which is discussed in section 3.3). By consistency condition (47) we must have $K \gg |a|$, hence the last term in (48) is small, though $K$ itself can still be large or small compared with unity. The solutions are thus approximately

$$\sigma_1^2 = -(K+4), \qquad \sigma_2^2 = -a\left(\frac{K_y}{K}\right)^2 \frac{K}{4+K}. \tag{49}$$

In the high wavenumber limit $K \gg 1$, this simplifies to

$$\sigma_1^2 = -K, \qquad \sigma_2^2 = -a\left(\frac{K_y}{K}\right)^2. \qquad (K \gg 1). \tag{50}$$

The first solution is a stable compressive wave, in which the restoring force is due to the perturbations in the curvature force. In terms of the dimensional quantities, the frequency of this wave, in the absence of rotation, is

$$\omega^2 = k\frac{B_z^2}{2\pi\Sigma}. \tag{51}$$

Since the restoring force is mediated by the potential field outside the disc, the wave frequency has the dependence $k^{1/2}$ which is characteristic also of the very analogous case of selfgravitating discs. The second solution is the interchange mode, unstable when $a < 0$, with the growth rate maximizing for $K_y \gg K_x$. It is stable for $a > 0$ with a behavior similar to internal gravity waves. For further discussion of the physics of these results, see ST. At wavenumbers $K \lesssim 1$, where the compressive wave frequency drops below the rotation frequency, rotation provides an additional restoring force in the compressive wave, and reduces the growth rate of the interchange mode without, however, changing the stability condition.

### 3.2 The incompressible limit

Like in the case of ordinary convection, interchange instability is described correctly, at large wavenumber, by the incompressible (Boussinesq) limit. As in the case of ordinary convection, this limit is obtained by taking $k \to \infty$ while keeping the 'pressure' force finite [the $\beta$-terms in (40-43)]. It has the advantage of showing the instability without interference from compressive waves. With $\hat{\boldsymbol{K}} = \boldsymbol{K}/K$ this limit gives

$$\hat{\boldsymbol{K}} \cdot \boldsymbol{U} \sim 1/K \quad (K \to \infty), \tag{52}$$

Hence $U_y \to -K_x/K_y U_x$. We multiply (40) by $K_y$, (41) by $K_x$ and subtract (thus effectively taking the curl of the equation of motion). Using (52) and (43) to eliminate $U_y$ and $F$ then yields

$$(1+s^2\tau^2)\frac{d^2 U_x}{d\tau^2} + 4\tau s^2 \frac{dU_x}{d\tau} + (2s^2+a)U_x = 0. \tag{53}$$

Hence the natural time scale in the incompressible case is just the shear time scale

$$v = \tau s. \tag{54}$$

In this time coordinate, and writing

$$f = (1 + v^2)U_x \tag{55}$$

this equation becomes

$$(1 + v^2)\frac{d^2 f}{dv^2} + \frac{a}{s^2}f = 0. \tag{56}$$

The equation is second order in time instead of fourth because by the incompressibility limit the compressive waves have been eliminated. The general solution of (56) can be written in terms of hypergeometric functions. Here we need mainly the asymptotic dependence of the solutions for large time, which are powers of $v$. The fastest growing solution has

$$U_x \sim v^n, \qquad n = [(1 - 4a/s^2)^{1/2} - 3]/2. \tag{57}$$

For $s = 0$ the index $n$ diverges, in accordance with the fact that in the uniformly rotating case the growth is exponential rather than algebraic. The condition that $U_x$ be a growing power of $v$ yields, with the definition of $a$

$$-N_m^2 > 2S^2. \tag{58}$$

The azimuthal velocity grows faster by one power of $v$, since by the incompressibility $U_y \sim K_x/K_y U_x$. Thus $U_y$ grows in the incompressible model whenever $N_m^2 < 0$, but only linearly or slower if (58) is not satisfied. It is a matter of debate whether such slow growth in $U_y$ only should be still called unstable. For uniform rotation, (58) reduces to the condition $-N_m^2 > 0$, in agreement with the results in ST. With (15),(44) condition (58) translates into

$$g_m \frac{d}{dr} \ln \frac{\Sigma}{B_z} > 2\left(r\frac{d\Omega}{dr}\right)^2. \tag{59}$$

If the length scale on which $\Sigma/B_z$ varies is of the order $r$, this means that the magnetic support of the disc must be comparable to the rotational support, before instability sets in. This implies that the magnetic energy of the system (disc plus external field) must be comparable to the gravitational binding energy. Only if the field strength varies with $r$ on a smaller length scale does interchange instability take place at smaller field strengths. For a disc with finite inner and outer radii, this condition also says that the intrinsic (nonrotating) instability time scale must be shorter than the shear time scale across the whole disc. It is probable that additional globally unstable modes would exist in such discs with boundaries.

We conclude that the incompressible limit predicts algebraic growth of interchange modes but only if the magnetic field is strong enough to start contributing to the radial support of the disc against gravity.

### 3.3 Numerical results

When the assumption of incompressibility is dropped, the full $4^{th}$ order system (40)-(43) must be integrated. Since the instability can be transient in the presence of shear, a growth rate or even an algebraic growth index is not necessarily a good measure of the effects of the instability. The growth of a small disturbance can be limited to a finite time interval, but the amplification factor over this interval can be quite large. This happens, for example, in the nonaxisymmetric form of the instability studied by Balbus and Hawley (1991).

For this reason, we measure the instability by the amplification factor of an initial disturbance over a finite interval in time. We determine this factor by integrating the linear equations (40-43) in time numerically. We start at $t = 0$ coresponding to $K_x = 0$ with the initial conditions

$$U_x = 1, \qquad U_y = 0, \qquad \beta = 0, \qquad F = 0. \tag{60}$$

In this way the perturbation starts as a purely incompressive ($\boldsymbol{K} \cdot \boldsymbol{U} = 0$) motion. Other choices yield a larger admixture of stable compressive waves, but very similar amplification factors, except at low amplification when the instability is rather marginal anyway. In Figure 1 the time development is shown, and compared with the incompressible limit discussed above. The incompressible results were obtained by integrating eq. (53) with the initial conditions

$$U_x = 1, \qquad dU_x/d\tau = 0. \tag{61}$$

These are the equivalent of (60) for the incompressible case. It is seen that the incompressible model is quite accurate when $K_y$ is large enough that condition (47) for consistency of the shearing sheet approximation is satisfied at all $t$. The asymptotic behaviour is accurately reproduced even when (47) is not satisfied initially; this is because for large $t$, the value of $K$ always becomes large enough that (47) holds. The comparison indicates additional, transient, growth around $t = 0$ for the smaller azimuthal wave numbers, in the compressible case. This may be a real effect of compressibility (some form of 'swing' amplification), but we can not establish this from the present model, since it happens only under conditions ($K(\tau) \lesssim |a|$) when the consistency condtion (47) is not satisfied.

We can conclude that the incompressible model seems to give the correct description of the instability whenever the wavenumber $K_y$ is large enough that the shearing sheet model applies.

Fig 2 shows the amplification factors after integration until $t = 10$, as functions of the driving parameter $a = N_m^2/\Omega^2$, the shear rate $s = S/\Omega$, and the dimensionless azimuthal wavenumber $K_y$.

The amplification factor shows two different forms of instability. At large $K_y$ we expect to find interchange instability essentially in its incompressible form, as discussed in section 3.2. This is shown in Figure 3a, for $K_y = 10$. The amplification factor show a broad hump centred at uniform rotation, $s = 0$. Amplification factors obtained from integrating equation (53) which is valid for the incompressible limit are shown for comparison in Fig. 3b. The agreement is good, in particular the incompressible instability condition (58) appears to be valid.

While this part of the instability behaves as expected, the amplification factor shows an additional form of instability at low $K_y$. An example is shown in Fig. 4, for $K_y = 0.25$. The growth appears to maximize at $K_y = 0$, the axisymmetric case. This is seen in Fig. 5, which show the amplification factor as a function of wavenumber $K_y$ and instability parameter $a$.

The axisymmetric case is easily treated analytically since in this case the radial wave number $K_x$ is time independent. From eqs (40-43) we have, with $K_y = 0$, the dispersion relation

$$\sigma^2 = -K_x - a - 2(2 + s). \tag{62}$$

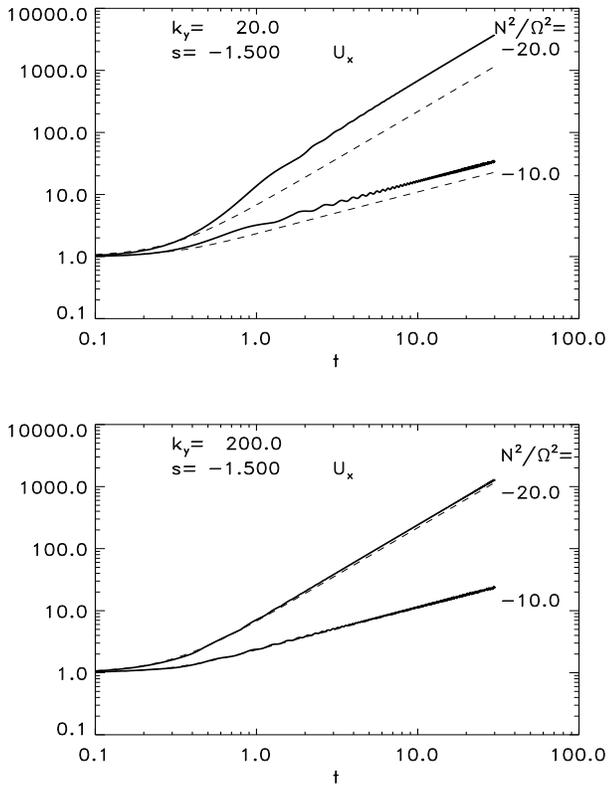

**Figure 1.** Velocity amplitude $U_x$ in interchange instability at $N_m^2/\Omega^2 = -20$ and $-10$, for $s = -1.5$, showing algebraic growth. Time in units of $\Omega^{-1}$. Dashed: growth predicted by the incompressible model. **top:** $K_y = 20$, **bottom:** $K_y = 200$,

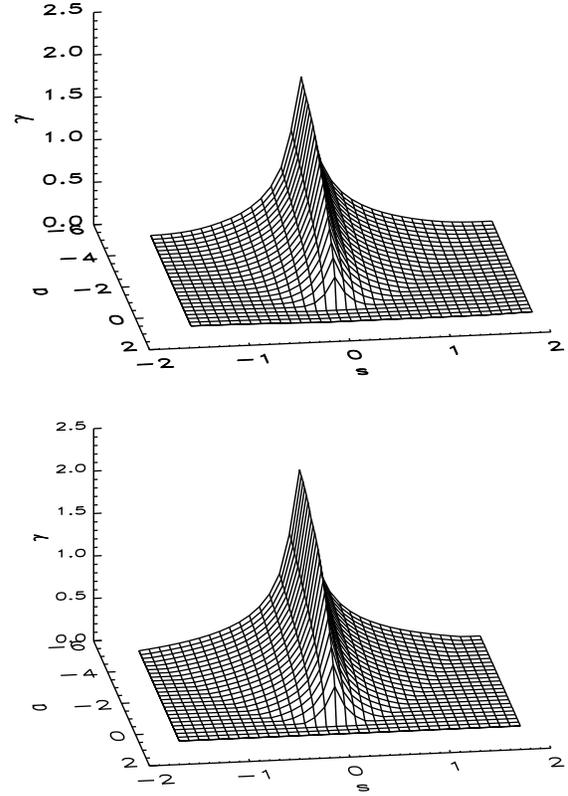

**Figure 3. a:** Amplification factor $A$ after $\Omega t = 30$ for $K_y = 10$ in terms of $\gamma = \ln A/30$, showing the stabilization of the interchange instability away from uniform rotation ($s = 0$). **b:** same derived analytically in the incompressible limit

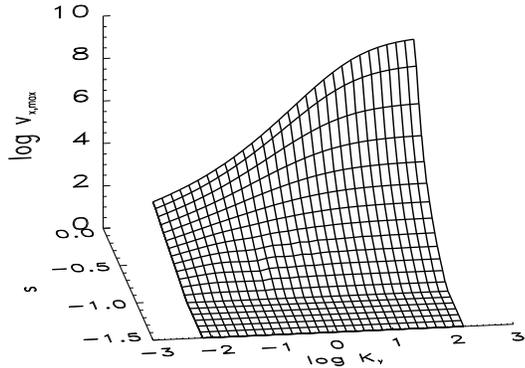

**Figure 2.** Amplification factors after $\Omega t = 10$ as functions of the dimensionless azimuthal wave number, for $a = N_m^2/\Omega^2 = -4 - 2s$.

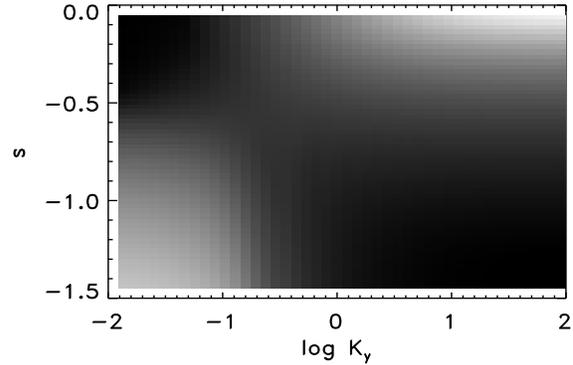

**Figure 4.** gray scale representation of the amplification factor (white highest) for $a = -3$.

In terms of the physical (dimensional) quantities, this predicts instability whenever

$$-N_m^2 > 2\frac{\Omega}{r}\frac{d}{dr}(\Omega r^2) + k_x \frac{B_z^2}{2\pi\Sigma}. \qquad (63)$$

The growth rate is of the order $\Omega$ except close to marginal stability, and instability first sets in at the lowest wavenumbers. The instability is therefore an intrinsically global one if it exists. This raises the question whether it can be treated correctly in the short wave approximation, on which (40-43) are based. Comparing (62) with (47) we see that the instability is excluded by condition (63) whenever the short wave approximation applies.

The last term in (63) is equal to $\omega_c^2$, where $\omega_c$ is the frequency of the compressive wave of wave number $k_x$ defined by (51). To see the physical interpretation of this term, consider the radial length scale $L_m$ over which the destabilizing field gradient exists (length scale of $N_m$). The prospective unstable wave must still fit into this length, for instability to occur, i.e. $k_x > 1/L_m$. Condition (63) then says that for axisymmetric instability to occur the intrinsic (nonrotating) growth must be faster than both the epicyclic frequency and the time for a compressive wave to cross the region with the destabilizing field gradient.

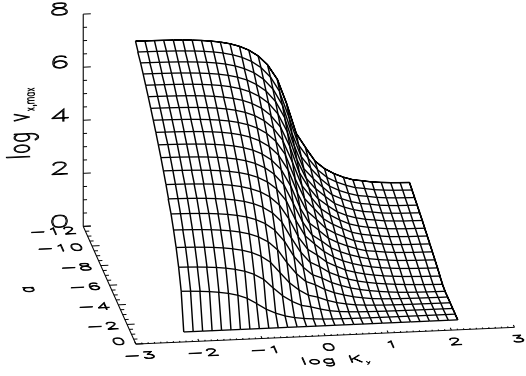

**Figure 5.** Amplification factors after $t = 5$ for $s = -3/2$

To study the long-wave form of instability properly, a more global form of stability analysis is needed. This is, in general, a cumbersome problem because of the nonlocal nature of the magnetic forces. A global result with which certain cases can be shown to be stable towards axisymmetric perturbations is given in the next section.

## 4 AXISYMMETRIC STABILITY

Analysis of the global stability problem is helped greatly by the indication, from the local analysis presented in the previous sections, that the low-wavenumber form of instability is essentially an axisymmetric one. This indicates that its mechanism, if it exists, must be simple. Assuming axisymmetry from the outset, the additional complication of transient instability due to shear disappears. We restrict ourselves to discs in which the polarity of $B_r$ does not depend on $r$. If the polarity of $B_z$ is also constant, this means that the sign of the magnetic acceleration $g_m$ is constant, but it is not necessary to assume this.

The equation of motion (1) can be written (in an inertial frame) as

$$\frac{d\boldsymbol{u}}{dt} = \frac{\boldsymbol{J} \times \boldsymbol{B}}{\Sigma} - \boldsymbol{g} \qquad (64)$$

where $\boldsymbol{J}(r,\phi)$ is the current integrated over the disc thickness, and $\boldsymbol{g}$ the gravity due to the central object. In the equilibrium state, this yields:

$$-\Omega^2 r = \frac{J_\phi B_z}{\Sigma} - \frac{GM}{r^2}, \qquad (65)$$

where the current is related to the radial field component by:

$$J_\phi = B_r^+ / 2\pi \qquad (66)$$

and $B_r^+$ is the radial field at the top surface of the disc, as in section 2.3. Because $B_r^+$ is of constant sign by assumption, and the stability of the field is unaffected by an overall change of sign of the field, we can take the signs of $B_r^+$ and $J_\phi$ to be positive. Equation (65) defines the equilibrium rotation rate $\Omega(r)$. Since we are considering an axisymmetric equilibrium with axisymmetric perturbations, the vertical component $B_z$ of the field can be written in terms of the vector potential $\boldsymbol{A}$ as

$$B_z = \frac{1}{r}\frac{\partial(rA_\phi)}{\partial r}, \qquad (67)$$

while the vector potential can be written in terms of the current distribution in the disc as

$$A_\phi = \mathcal{O}(J_\phi), \qquad (68)$$

where the self-adjoint operator $\mathcal{O}$ is defined by

$$\mathcal{O}(J_\phi) = \frac{1}{4\pi}\int_s ds\, \frac{\cos(\phi-\phi')\,J_\phi(r',\phi')}{[r^2 + r'^2 - 2rr'\cos(\phi-\phi')]^{1/2}}. \qquad (69)$$

The integral is taken over the surface $s$ of the disc. The operator has the property that $\mathcal{O}(F)$ is positive definite if $F$ is positive definite, hence $A_\phi$ is positive definite.

Axisymmetric perturbations are not sheared, and their linear development can be described in terms of normal modes. Linearizing (64), setting $\partial/\partial t = \gamma$, the radial and azimuthal components can be combined to yield

$$\gamma^2 \xi_r = -\kappa^2 \xi_r - J_\phi \xi_r \frac{d}{dr}\left(\frac{B_z}{\Sigma}\right) + \frac{J_\phi' B_z}{\Sigma}, \qquad (70)$$

where $\boldsymbol{\xi}$ is the displacement ($\boldsymbol{u} = \partial\boldsymbol{\xi}/\partial t$), a prime denotes a perturbed quantity and the epicyclic frequency $\kappa$ is given by

$$\kappa^2 = 2\Omega \frac{1}{r}\frac{d}{dr}(\Omega r^2). \qquad (71)$$

¿From the induction equation $\partial \boldsymbol{B}/\partial t = \nabla \times (\boldsymbol{u} \times \boldsymbol{B})$ we have, using axisymmetry,

$$A_\phi' = -\xi_r B_z = \mathcal{O}(J_\phi') \qquad (72)$$

Thus $J_\phi'$ can be written in terms of $\xi_r$, so that (70) can be written as

$$\gamma^2 \xi_r = \mathcal{L}(\xi_r), \qquad (73)$$

where $\mathcal{L}$ is the self-adjoint operator defined by the RHS of (70). Multiplying (70) by $\Sigma \xi_r^*$ and integrating over $s$ we have

$$\int_s ds\, \Sigma \xi_r^* \mathcal{L}(\xi_r) = \int_s ds\, \Sigma |\xi_r|^2 [-\kappa^2 - J_\phi \frac{d}{dr}(\frac{B_z}{\Sigma})] -$$
$$\int_s ds\, J_\phi'^* \mathcal{O}(J_\phi') \equiv -W. \qquad (74)$$

For stability, it is sufficient that the energy $W$ be positive definite. The main problem is the estimation of the magnetic energy term

$$\int_s ds\, J_\phi'^* \mathcal{O}(J_\phi') \qquad (75)$$

which is positive and hence stabilizing. It corresponds to the restoring force that develops (via the potential field above the disc) when the field lines threading the disc are compressed together or expanded. For nonaxisymmetric perturbations, the energy expenditure in such compressions/expansions can be minimized by taking rotational displacements (i.e. interchanges), but this is not possible for axisymmetric perturbations. To find a sufficient condition for stability, we need to find a sufficiently large minimum value for (75). To do this, consider the variational problem of finding the minimum of $\mu$ defined by

$$\mu = \frac{\int_s ds\, J_\phi'^* \mathcal{O}(J_\phi')}{\int_s ds\, f(r)|A_\phi'|^2} = \frac{\int_s ds\, A_\phi'^* \mathcal{O}^{-1}(A_\phi')}{\int_s ds\, f(r)|A_\phi'|^2}. \tag{76}$$

for some positive definite $f(r)$. The minimum of $\mu$ is the smallest eigenvalue of the equation

$$\mathcal{O}^{-1}(A_\phi') = \mu f A_\phi'. \tag{77}$$

If we choose

$$f(r) = J_\phi/A_\phi \tag{78}$$

(which is positive definite since $J_\phi$ and $A_\phi$ are), this equation can be written as

$$A_\phi \mathcal{O}^{-1}(A_\phi') = \mu A_\phi' \mathcal{O}^{-1}(A_\phi). \tag{79}$$

It is apparent that (79) has the solution $A_\phi' = A_\phi$ with $\mu = 1$ and the corresponding $J_\phi' = J_\phi$. The current density for this eigenfunction is everywhere positive as it must be for the smallest eigenvalue $\mu$ which is therefore unity. Hence the variational problem (76) yields

$$\int_s ds\, J_\phi'^* \mathcal{O}(J_\phi') \geq \int_s ds\, f(r)|A_\phi'|^2. \tag{80}$$

The energy then satisfies:

$$W \geq \int_s ds\, \Sigma |\xi_r|^2 [\kappa^2 + J_\phi \frac{d}{dr}\left(\frac{B_z}{\Sigma}\right) + \frac{J_\phi B_z^2}{A_\phi \Sigma}]. \tag{81}$$

Using the definition of $\kappa$ (71) and the equilibrium condition (65), while writing $J_\phi$ in terms of $B_r$ and $A_\phi$ in terms of $B_z$ by inversion of (67), this can be written as

$$W \geq \int_s ds\, \Sigma |\xi|^2 \{\Omega_K^2 + g_m \left[-\frac{1}{r^3 B_r}\frac{d}{dr}(r^3 B_r) + \frac{r B_z}{\int_0^r r B_z dr}\right]\}, \tag{82}$$

where $\Omega_K = (GM/r^3)^{1/2}$ is the kepler rotation rate and the magnetic acceleration $g_m$ is given by (14). It follows that stability is guaranteed if everywhere in the disc we have

$$g_m \left[r\frac{d}{dr}\ln(r^3 B_r) - \frac{r^2 B_z}{\int_0^r r B_z dr}\right] \leq \Omega_K^2 r. \tag{83}$$

This may be compared with the estimate (63) from the shearing-sheet analysis, which predicts stability against axisymmetric perturbations if

$$g_m \left[r\frac{d}{dr}\ln(r^3 B_r) - |\frac{B_z}{B_r}k_x r|\right] \leq \Omega_K^2 r. \tag{84}$$

The term in $k_x$ corresponds to the stabilizing magnetic energy term (second term) in (83). While the shearing sheet result suggests that the stabilizing effect of the magnetic energy term can be removed by taking a sufficiently low wave number, the result (83) shows that in fact this term has a finite minimum value which can not be removed for any wave number.

As an example consider a disc in which the variables in the equilibrium state vary as powers of $r$, in such a way that

$$B_z(r, z = 0) \sim r^{-\nu}. \tag{85}$$

The potential field corresponding to this $B_z$ has a radial component $B_r(r, z = 0) \sim r^{-\nu}$ (cf. Sakurai, 1987). Thus, the angle of inclination of the field lines at the disc surface is constant. For such a field, sufficient condition (83) reduces to

$$\Omega_K^2 - g_m/r = \Omega^2 > 0, \tag{86}$$

independent of $\nu$. In equilibrium, the magnetic support $g_m$ can not exceed the gravity $\Omega_K^2 r$, so that all equilibria which vary as powers of $r$ are stable to axisymmetric displacements.

To prove stability to axisymmetric perturbations for all equilibria in which $J_\phi$ has constant sign, a stronger version of (83) would be needed. This is because the sufficient condition can be defeated near places where $B_r$ has steep gradients, by choosing a $\xi_r$ that peaks near this gradient. As the axisymmetric estimate (84) shows, such smaller scale perturbations are stabilized more strongly than suggested by (83).

## 5  CONCLUSIONS AND DISCUSSION

A thin disc threaded by a poloidal field which provides part of the support against gravity has two kinds of magnetic wave modes for motions in the plane of the disc. A compressive wave is present in which the restoring force is due to the potential field outside the disc. It has been studied before by ST and Tagger et al. (1990). A radial gradient in the field strength introduces modes analogous to internal gravity waves. These can become unstable in a sufficiently steep gradient in the field strength, and then behave as expected from convective or magnetic interchange modes. We find that these interchange instabilities in a shearing disc behave essentially as expected from an incompressible approximation (analogous to the Boussinesq approximation in convection) which becomes exact for short wavelengths.

At longer wavelengths, indications are found for a second form of instability. Its growth maximizes for axisymmetric perturbations. In all cases however, it appears only when the wavelengths are longer than allowed by the local (shearing sheet) approximation under which the results were obtained. This form of instability, discussed in some detail in Lubow and Spruit (1995), is therefore possibly an artefact. A simple global stability bound derived in section 4 also suggests stability for axisymmetric perturbations. Better global stability calculations are needed to settle this issue however.

In previous work on uniformly rotating discs (ST), the instability was found to grow on the rather short magnetic time scale $g_m/r$ where $g_m$ is the radial acceleration due to the magnetic field. The condition for instability was found to be that the surface density $\Sigma$ and magnetic field strength $B_z$ vary such that

$$\frac{d}{dr}(\frac{\Sigma}{B_z}) > 0. \tag{87}$$

The stabilizing effect of the shear flow in the discs studied here turns out to be dramatic. We find that significant growth of perturbations is possibe only if the growth rate in a corresponding system without shear is of the order of the dynamical time. The incompressible model gives, as the condition for significant growth of linear perturbations:

$$g_m \frac{d}{dr}(\frac{\Sigma}{B_z}) > 2S^2, \tag{88}$$

where $S$ is the shear rate $rd\Omega/dr$. The instability happens on short length scales and has algebraic (power law) growth. For a disc with finite inner and outer radii, this condition also says that the intrinsic (nonrotating) instability time scale must be shorter than the shear time scale across the whole disc. It is probable that additional globally unstable modes would exist in such discs with boundaries.

Integrations without making the incompressible approximation agree closely with condition (88). If we assume $S \approx \Omega$, denote the length scale on which $\Sigma/B_z$ varies by $L$, and the disc thickness by $H$, (88) can be written as

$$v_A^2/c_s^2 \gtrsim L/H, \quad \text{or} \quad \beta \lesssim H/L \tag{89}$$

where $v_A$ is the Alfvén speed in the disc, $c_s \approx \Omega H$ the sound speed, and $\beta$ the plasma-beta. If $\Sigma/B_z$ varies smoothly, i.e. $L \sim r$, instability requires the magnetic energy density inside the disc to exceed the geometric mean of the thermal and kinetic (orbital) energy densities. If the energy of the external magnetic field is included, the condition is equivalent to the requirement that the magnetic energy of the system (disc plus field) must be comparable to the orbital kinetic energy. This happens when the magnetic curvature force also provides a significant fraction of the support against gravity.

If, on the other hand, the length scale of the unstable gradient is $H$ (the smallest meaningful length scale for this gradient), instability requires $v_A \gtrsim c_s$ or $\beta \lesssim 1$. This is just the condition at which instability of poloidal fields by the Velikhov-Chandrasekhar mechanism (Balbus and Hawley, 1991) stops operating. Thus it appears that there is a significant region in field strengths, $1 \lesssim v_A/c_s \lesssim r/H$, where smoothly varying fields are stable both to interchanges and Velikhov-Chandrasekhar-Balbus-Hawley type instabilities.

Discs in which a poloidal magnetic field provides part of the support against gravity are a natural consequence of star formation in a magnetic environment. An initially weak poloidal field is concentrated by the accretion flow, and its energy density will eventually become a significant force opposing accretion, if the field is frozen into the flow. In this case, the degree of support is limited by the tendency of magnetic fields towards interchange instability, which effectively allows the field to diffuse radially through the disk. Though ambipolar diffusion also has this effect, and may be able to leak fields out of the disc during most of its history, interchange of some form may well be the main mechanism limiting the advection of a poloidal field in the hot inner regions of the disc. Our results indicate that quite strong poloidal fields may be expected in these regions. Since such fields are the main ingredient in the magnetic acceleration mechanism of jets and winds, our result lends support to the idea that strong poloidal fields are responsible for these flows.

### ACKNOWLEDGMENTS
The work reported here was done as part of the research network 'accretion onto compact objects and protostars' supported by EC grant ERB–CHRX–CT93–0329. R.S. has been supported by 'Deutscher Akademischer Austauschdienst im Rahmen des 2. Hochschulsonderprogramms'.